\documentclass[twocolumn,aps,superscriptaddress,showpacs,nofootinbib,floatfix]{revtex4}
\usepackage{epsfig,bm,feynmf}
\usepackage{graphics}
\usepackage{amsmath}
\usepackage[normalem]{ulem}  
\usepackage[dvips]{color} 

\newcommand{\ans}[1]{{\sf\color[rgb]{0,1,0}{#1}}}
\renewcommand\sout{\bgroup \color{red} \ULdepth=-.5ex \ULset}
\begin{document}


\title{Discrepancy in low transverse momentum dileptons from relativistic heavy-ion collisions}


\author{Taesoo Song}\email{taesoo.song@theo.physik.uni-giessen.de}
\affiliation{Institut f\"{u}r Theoretische Physik, Universit\"{a}t
Gie\ss en, Germany}

\author{Wolfgang Cassing}
\affiliation{Institut f\"{u}r Theoretische Physik, Universit\"{a}t
Gie\ss en, Germany}

\author{Pierre Moreau}
\affiliation{Institute for Theoretical Physics, Johann Wolfgang
Goethe Universit\"{a}t, Frankfurt am Main, Germany}

\author{Elena Bratkovskaya}
\affiliation{Institute for Theoretical Physics, Johann Wolfgang
Goethe Universit\"{a}t, Frankfurt am Main, Germany}
\affiliation{GSI
Helmholtzzentrum f\"{u}r Schwerionenforschung GmbH, Planckstrasse 1,
64291 Darmstadt, Germany}


\begin{abstract}
The dilepton transverse momentum spectra and invariant mass spectra
for low $p_T <0.15$~GeV/c in Au+Au collisions of different centralities
at $\sqrt{s_{NN}}$ = 200 GeV are studied within the parton-hadron-string dynamics (PHSD)
transport approach. The PHSD describes the whole evolution of the system on a
microscopic basis, incorporates  hadronic and partonic degrees-of-freedom,
the dynamical hadronization of partons and hadronic rescattering.
For dilepton production in p+p,  p+A and A+A reactions the PHSD incorporates the leading
hadronic and partonic channels (also for heavy flavors) and includes in-medium effects
such as a broadening of the vector meson spectral functions in hadronic matter and
a modification of initial heavy-flavor correlations by interactions
with the partonic and hadronic medium.
The transport calculations reproduce well the momentum integrated invariant mass spectra from the STAR Collaboration
for minimum bias Au+Au collisions at $\sqrt{s_{NN}}$ = 200 GeV, while the description
of the STAR data - when gating on low $p_T < 0.15$ GeV/c - is getting  worse
when going from central to peripheral collisions.
An analysis of the transverse momentum spectra shows that the data for peripheral
(60-80\%) collisions are well reproduced for $p_T>0.2$ GeV/c while the strong
peak at low $p_T < 0.15$ GeV/c, that shows up in the experimental data for the
mass bins ($0.4 < M < 0.7$ GeV and $1.2 < M < 2.6$ GeV), is fully missed
by the PHSD and cannot be explained by the standard in-medium effects.
This provides a new puzzle for microscopic descriptions of low $p_T$ dilepton data from the STAR
Collaboration.
\end{abstract}


\maketitle


Relativistic heavy-ion collisions are well suited to produce hot and
dense matter in the laboratory. Whereas low-energy collisions
create nuclear matter at high baryon chemical potential and
moderate temperature, high-energy collisions  at the Relativistic
Heavy-Ion Collider (RHIC) or the Large Hadron Collider (LHC) produce
a dominantly partonic matter at high temperature and almost
vanishing baryon chemical potential. The latter    
is controlled by lattice quantum chromodynamics (lQCD) which
shows that the phase transition between the quark-gluon plasma (QGP)
and the hadronic system is a crossover at low baryon chemical
potential~\cite{Bernard:2004je,Aoki:2006we,Bazavov:2011nk}.

Since the partonic matter in relativistic heavy-ion collisions
survives  only for a couple of fm/c within a finite volume, it is
quite challenging to investigate its properties. In this context
hard probes (heavy flavor or jets) and penetrating probes (photons
or dileptons) are of particular interest. Dileptons have the
advantage of an additional degree of freedom compared to
photons, i.e. their invariant mass, which allows to roughly separate
hadronic and partonic contributions by appropriate mass cuts \cite{rapp5}. For
example, dileptons with invariant mass  less than 1.2 GeV dominantly
stem from hadronic decays while those with  invariant masses between
1.2 GeV  and 3 GeV stem from partonic interactions and correlated
semileptonic decays of heavy flavor hadrons. In the first case it is
possible to study the modification of hadron properties such as a
$\rho$ meson broadening or a mass shift in nuclear matter~
\cite{ChSym,PHSDreview}. On the other hand the dileptons with
intermediate masses provide information on the properties of
partonic matter once the background from semileptonic heavy flavor
decays is subtracted. This background overshines the partonic
contribution at RHIC and LHC energies and is subleading only at
collision energies per nucleon below about $\sqrt{s_{NN}}$ = 10
GeV~\cite{Song:2018xca}.

Recently, dielectrons in Au+Au collisions at $\sqrt{s_{\rm NN}}=$
200 GeV  have been measured as a function of transverse
momentum~\cite{yang2018} for different centralities. It turned out
as a surprise that the yield of dielectrons is largely enhanced at
low transverse momentum - compared to expected hadronic decays - in
particular in peripheral collisions of 60-80\% centrality.
%
%
In case the low $p_T$ peak would be measured in ultra-peripheral
collisions \cite{Adams:2004rz} -
for impact parameters larger than roughly twice the radius of the nuclei -
one could attribute it to a coherent source from the strong electromagnetic
fields generated by the charged spectators \cite{yang2018}.
However, an interesting point is that the low $p_T$ enhancement is observed
in peripheral collisions with dominant hadronic reaction channels,
which are expected to be under control by independent $p+p$ measurements.
This raises severe doubts on a coherent nature of the observed phenomenon.
These surprising observations come up as a puzzle and in this work we will
investigate the question if hadronic and partonic in-medium effects might
be the	 origin for  the anomalous enhancement of dielectrons at low
transverse momentum in peripheral collisions.

We will employ the microscopic parton-hadron-string dynamics (PHSD)
transport approach where quarks and gluons in the quark-gluon plasma
are off-shell massive strongly interacting quasi-particles. The
masses of quarks and gluons are assigned from their spectral
functions at finite temperature whose pole positions and widths are,
respectively, given by the real and imaginary parts of partonic
self-energies~\cite{PHSDreview}. The PHSD approach has successfully
described experimental data in relativistic heavy-ion collisions for
a wide range of collision energies from the SchwerIonen-Synchrotron (SIS)
to the LHC for many hadronic as well as
electromagnetic observables \cite{PHSDreview,Volo,Eduard}.



The production channels for dileptons in relativistic heavy-ion
collisions  may be separated into three different classes: i)
hadronic production channels, ii) partonic production channels and
iii) the contribution from the semileptonic decay of heavy-flavor
pairs. The production of dileptons in the hadronic phase includes
the following steps: First a resonance $R$ is produced either in a
nucleon-nucleon (NN) or meson-nucleon (mN) collisions.
 The produced resonance $R$ may produce dileptons directly through Dalitz decay,
for example, $\Delta \to e^+e^-N$, or the resonance $R$ decays to a meson
which produces dielectrons through direct decay ($\rho, \omega,
\phi$) or  Dalitz decay ($\pi^0, \eta, \omega$).
Additionally the resonance $R$ may decay to another resonance
$R^\prime$  which then produces dileptons through Dalitz decay. In
the PHSD we take into account also dilepton production by two-body
scattering such as $\pi+\rho$, $\pi+\omega$, $\rho+\rho$, $\pi+a_1$
\cite{Olenasps}, although the contributions are
subleading. An important point is the modification of the
vector-meson spectral functions ($\rho, \omega, \phi$), i.e. the
collisional broadening of the vector-meson widths in nuclear matter
is incorporated in PHSD (by default)\cite{Brat08dil}, which leads to results
consistent with the experimental data on dileptons from
SIS to LHC energies~\cite{Brat08dil,Olenasps,PHSDreview}.

In partonic matter dileptons are produced through the  channels
$q\bar q\to\gamma^*$, $q\bar q\to\gamma^*g$ and $qg\to\gamma^*q$
($\bar q g\to\gamma^* \bar q$) where the virtual photon $\gamma ^*$
decays into $e^+e^-$ or $\mu^+ \mu^-$ pair. We note that
$q(\bar{q})$ and $g$ in the above processes stand for off-shell
partons and the effective propagators for quarks and gluons from the
Dynamical QuasiParticle Model (DQPM) \cite{PHSDreview} have been employed for the
calculation of the differential cross sections in
Refs.~\cite{Song:2018xca,olena2010}. We recall that the dileptons
from the QGP are produced in the early stage of heavy-ion collisions
and have a relatively large invariant mass and high effective
temperature.

The production of dileptons from heavy-flavor pairs is different
from the other channels  since the lepton and anti-lepton are
produced in separate semi-leptonic decays. However, since heavy
flavor is always produced by pairs, it contributes to dilepton
production with the probability that both heavy flavor and
anti-heavy flavor have semi-leptonic decays. Furthermore, the heavy
flavor pairs - produced very early in heavy-ion collisions - suffer
from strong interactions with the partonic or hadronic medium and
thus the kinematics of the pair change in time. E.g., the
heavy flavor quarks are suppressed at high transverse momentum due
to the energy loss in partonic  matter, while slow heavy flavor
quarks are shifted to larger momenta due to collective flow. These
modifications of heavy flavor pairs in heavy-ion collisions affect
the spectrum of dileptons as demonstrated in Ref.
\cite{Song:2018xca}.

\begin{figure} [h]
\centerline{
\includegraphics[width=8.6 cm]{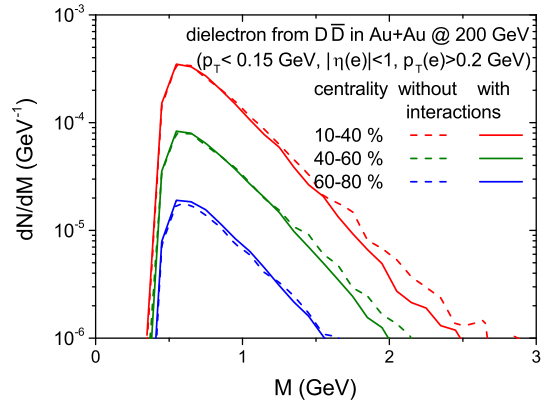}}
\caption{Invariant mass spectra of dileptons from $D\bar{D}$ pairs
with and without partonic and hadronic interactions in 10-40, 40-60,
and 60-80 \% central Au+Au collisions at $\sqrt{s_{\rm NN}}=$ 200
GeV.} \label{ddbar}
\end{figure}

To show these effects quantitatively we compare in Fig. \ref{ddbar}
the invariant mass spectra of dileptons with transverse  momenta
less than 0.15 GeV/c from heavy-flavor pairs with and without partonic
and hadronic interactions in 10-40, 40-60, and 60-80 \% central
Au+Au collisions at $\sqrt{s_{\rm NN}}=$ 200 GeV. The figure shows
that the interactions of heavy flavors soften the invariant mass
spectra of dileptons especially in central collisions while the
effect is hardly visible in very peripheral collisions. This change
in slope is due to energy loss for high momentum heavy flavors by
interactions which randomizes  the correlation angle between charm
and anti-charm quarks \cite{Song:2018xca}. The softening of the mass
spectrum becomes weaker with decreasing centrality since there are
less and less interactions of charm quarks.

Summarizing, there are three different medium modifications on
dileptons  in relativistic heavy-ion collisions, which cannot be
described by hadronic cocktails: 1) the broadening of the
vector-meson spectral functions in nuclear matter, 2) dilepton
production from partonic interactions, and 3) the modification of
the dilepton spectra from heavy-flavor pairs due to strong charm or
beauty scatterings in particular in the partonic phase. We will now
explore these effects on the momentum and mass spectra for dileptons
in Au+Au collisions at $\sqrt{s_{\rm NN}}=$ 200 GeV for different
centrality classes.


Since dileptons have the invariant mass as an additional degree of
freedom, compared to photons or other hadronic probes in heavy-ion
collisions, it potentially provides more information on the matter produced in
these collisions.

\begin{figure} [h]
\centerline{
\includegraphics[width=8.6 cm]{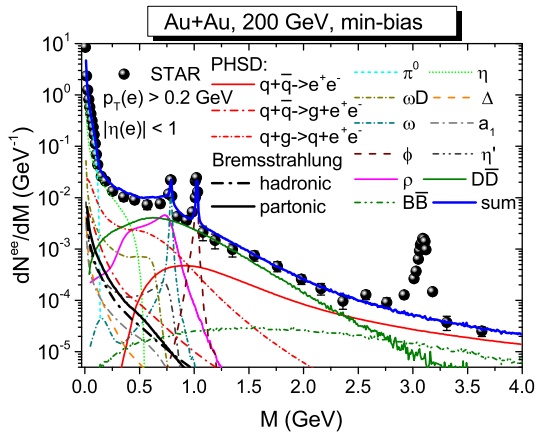}}
\caption{Invariant mass spectrum of dileptons from the PHSD in
minimum-bias Au+Au collisions at $\sqrt{s_{\rm NN}}=$ 200 GeV in
comparison to the experimental data from the STAR
collaboration~\cite{Adamczyk:2015lme}. The different channels are specified in the legend.} \label{rhic}
\end{figure}

This is demonstrated in Fig. \ref{rhic} where  the invariant mass
spectrum  of dielectrons in minimum-bias Au+Au collisions at
$\sqrt{s_{\rm NN}}=$ 200 GeV is shown for the constraint that the
transverse momentum of electron and that of position both are larger
than 0.2 GeV/c and each rapidity is smaller than unity, i.e. $|y_e|
\leq 1$.
We note that dielectron Bremsstrahlung from both partonic and hadronic collisions - as suggested long ago
\cite{add1,add2,add3,add4,add5,add6} - is added to our previous study~\cite{Song:2018xca}, although the contributions are subleading. For an estimate of the order of magnitude the differential Bremsstrahlung cross section is evaluated in the soft-photon approximation:
\begin{eqnarray}
E\frac{d^2\sigma(e^+e^-)}{dMd^3 p}=\frac{\alpha^2}{6\pi^3M}\frac{|\epsilon\cdot J|^2}{e^2} \frac{\lambda^{1/2}(s_2,m_3,m_4)}{\lambda^{1/2}(s,m_3,m_4)}\sigma_{el},
\end{eqnarray}
where $\epsilon_\mu$ is the polarization vector of the virtual photon and $J_\mu$ the electromagnetic current of the incoming and outgoing particles in the reaction $1+2\rightarrow 3+4+\gamma^*$. Furthermore, $\sigma_{el}$ is the elastic scattering cross section and $\lambda^{1/2}(s_2,m_3,m_4)$ is the three momentum of particle 3 or 4 in their center-of-mass frame at the invariant energy $s_2=(p_3+p_4)^2$~\cite{PHSDreview}.
One can see from Fig. \ref{rhic} that many hadronic sources contribute to the
low-mass dilepton spectrum while the intermediate-mass range is
dominated by the contribution from heavy-flavor pairs and that from
partonic interactions. We note that the $\rho$ meson considerably
broadens and that the contribution from charmonia is not included in
the PHSD calculations which explains the missing peak in the data
from the STAR collaboration~\cite{Adamczyk:2015lme} at about 3.1 GeV
of invariant mass. Nevertheless, the description of the inclusive
dilepton spectra within PHSD is very good for lower invariant masses.

In view of the completely different contributions for low-mass
dileptons and for intermediate-mass dileptons, it is helpful to
separate them for studying transverse-momentum spectra.

\begin{figure} [h]
\centerline{
\includegraphics[width=8.6 cm]{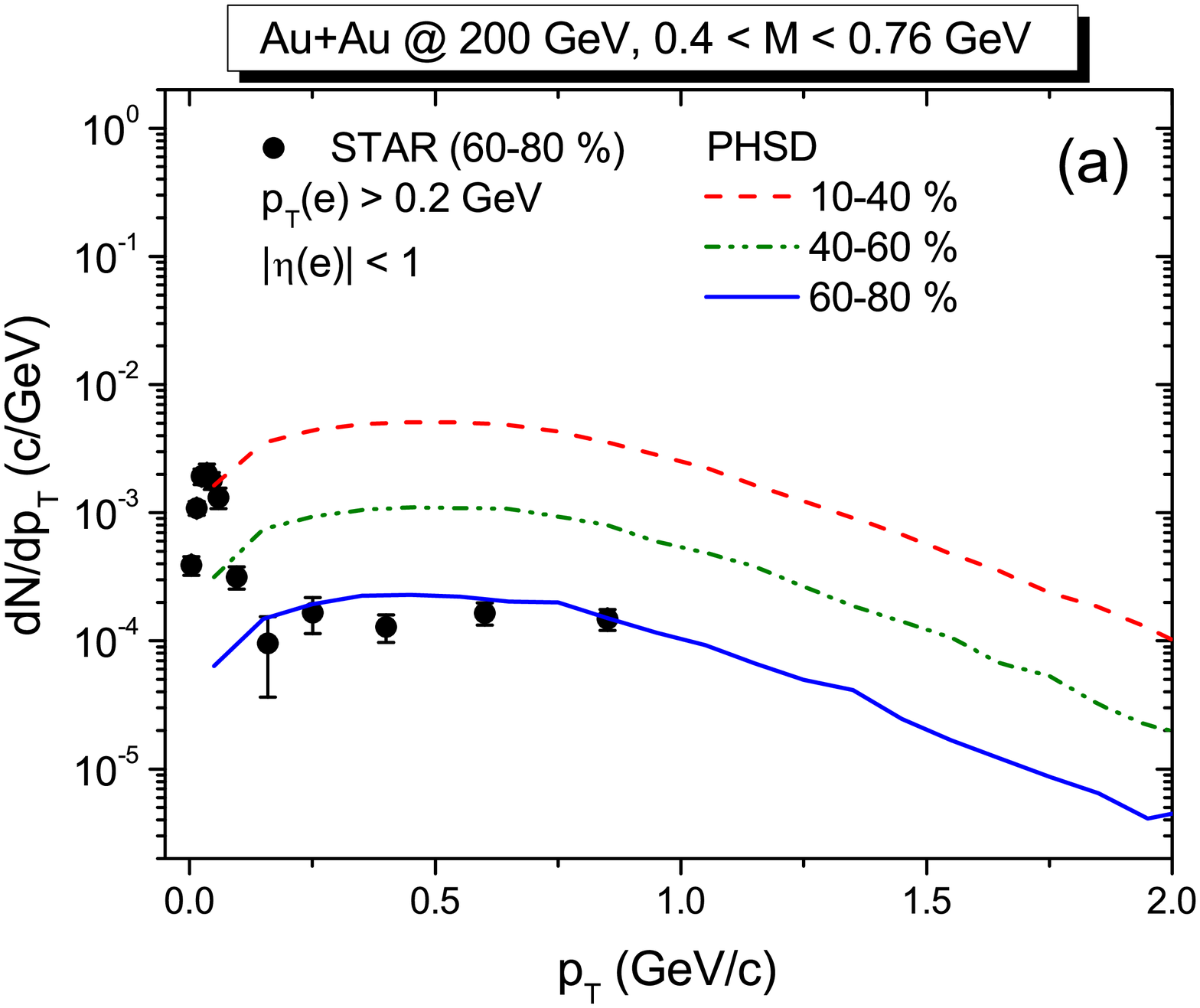}}
\centerline{
\includegraphics[width=8.6 cm]{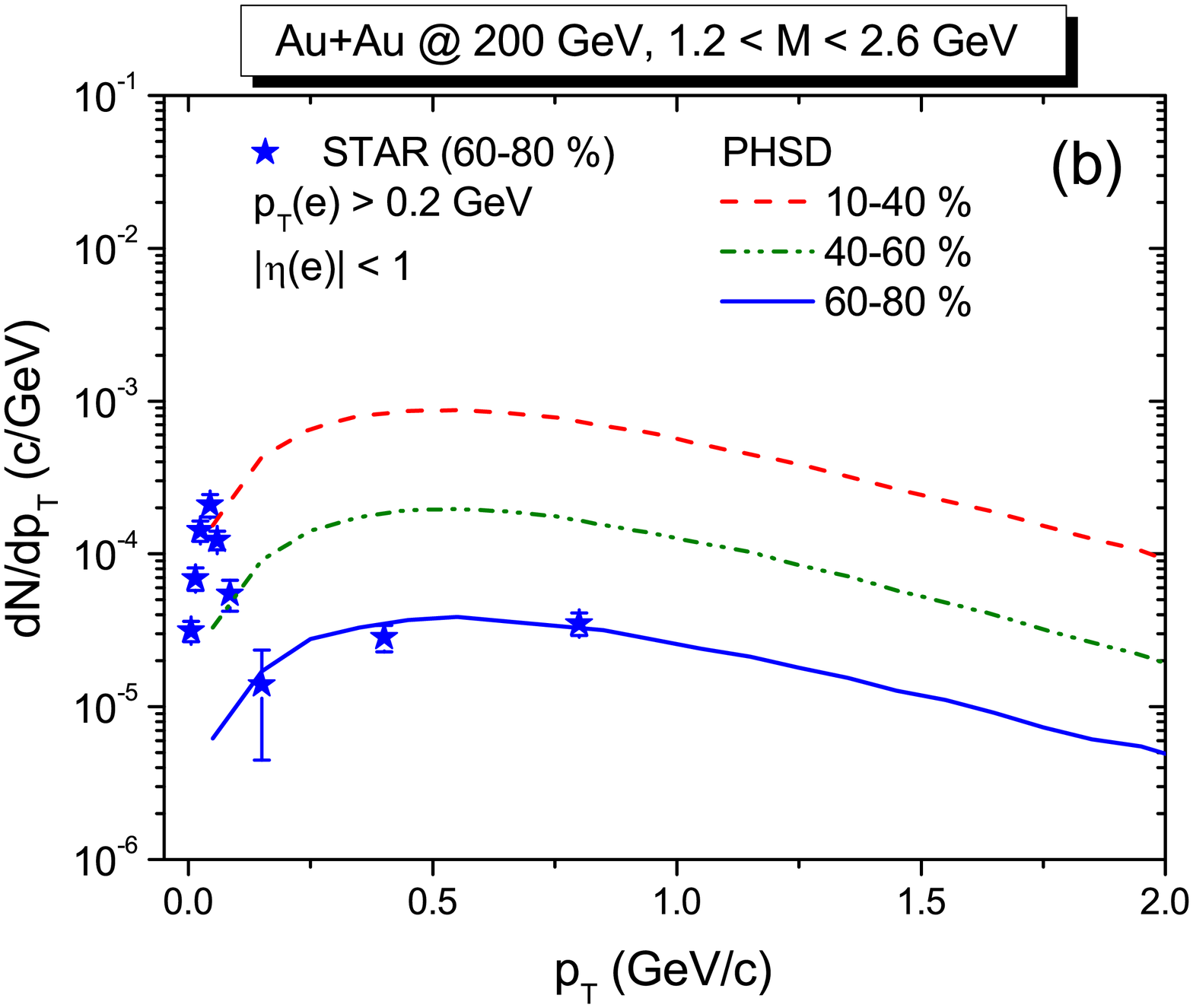}}
\caption{Transverse momentum spectra of (a) low-mass and (b)
intermediate-mass dileptons in 10-40, 40-60, and 60-80 \% central
Au+Au collisions at $\sqrt{s_{\rm NN}}=$ 200 GeV in comparison to
the experimental data (for 60-80 \% central
collisions) \cite{yang2018}.} \label{spec1}
\end{figure}

We show in Fig. \ref{spec1} the transverse momentum spectra of
low-mass  ($0.4 < M < 0.79$ GeV) and intermediate mass ($1.2 < M <
2.6$ GeV) dileptons for the same acceptance cuts as in Fig.
\ref{rhic} for 10-40, 40-60, and 60-80 \% central Au+Au collisions
at $\sqrt{s_{\rm NN}}=$ 200 GeV. The yields of low-mass dileptons
within the acceptance cuts are $4.85 \times 10^{-3}$, $1.05 \times
10^{-3}$, and $2.1 \times 10^{-4}$ for 10-40, 40-60, and 60-80 \%
central collisions, respectively. For the intermediate-mass
dileptons they are, respectively, $1.0 \times 10^{-3}$, $2.2 \times
10^{-4}$, and $4.5 \times 10^{-5}$. Comparing the low-mass and
intermediate-mass dileptons, the ratio of the dilepton yields in
10-40 \% central collisions to that in 40-60 \% or 60-80 \% central
collisions is very similar. This demonstrates that the dependence of
the dilepton yield on invariant mass is not so sensitive to the
centrality in heavy-ion collisions, if the collision energy is the same.
The shape of the transverse momentum spectra of dileptons is neither
sensitive to the centrality as shown in Fig. \ref{spec1}. However,
with increasing transverse momentum the spectrum of low-mass
dielectron decreases faster than that of intermediate-mass dileptons as
expected.

Furthermore, in Fig. \ref{spec1} we compare the results from the
PHSD with the experimental data for 60-80 \% central collisions. It
is seen that the PHSD reproduces very well the experimental spectra
both for low-mass dileptons and for intermediate-mass dileptons down
to $p_T \approx$ 0.15 GeV/c. The experimental data show an anomalous
enhancement of dileptons below $p_T \approx$ 0.15 GeV/c for the very
peripheral collisions, which is not described by the PHSD at all.

\begin{figure} [h]
\centerline{
\includegraphics[width=8.6 cm]{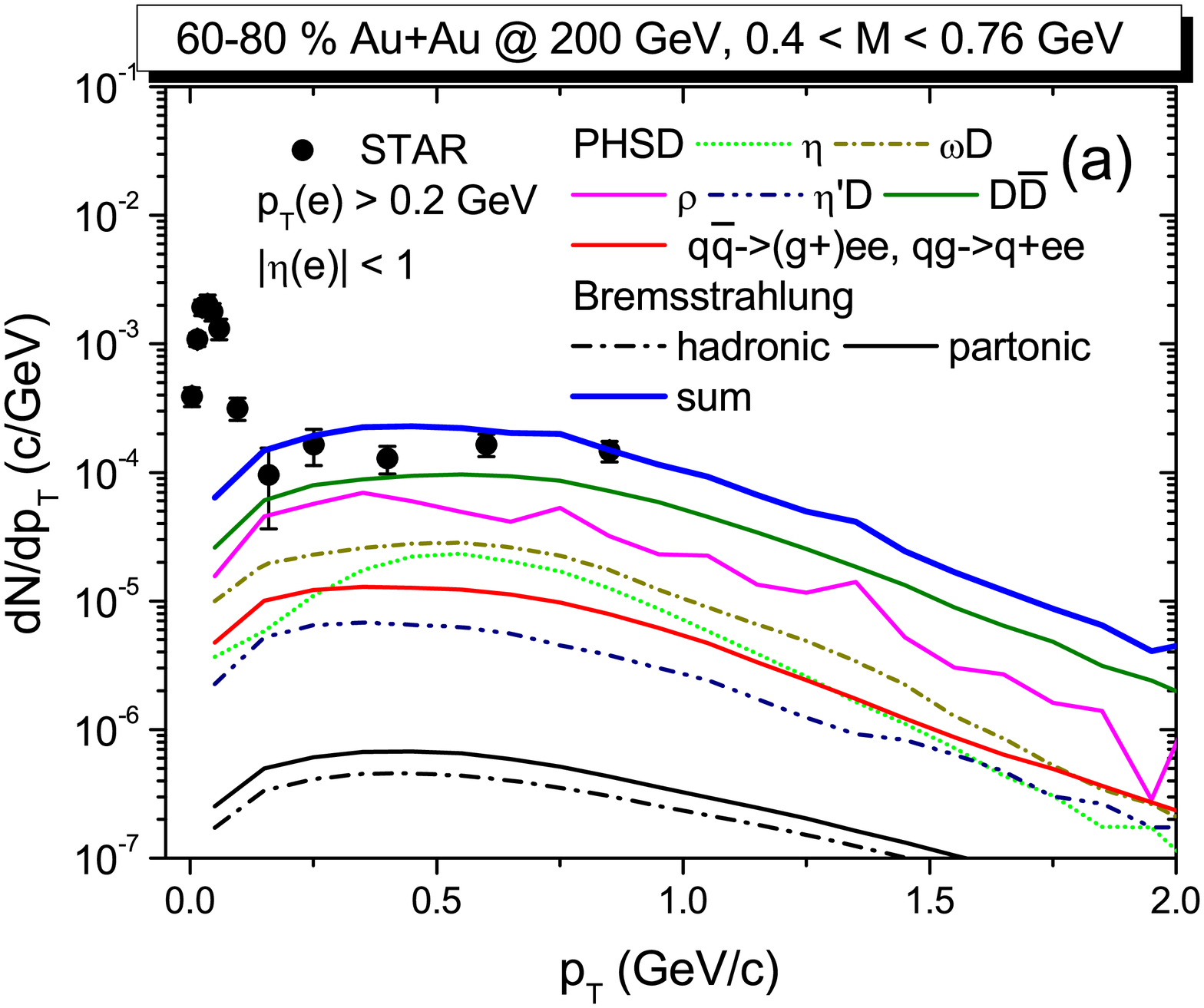}}
\centerline{
\includegraphics[width=8.6 cm]{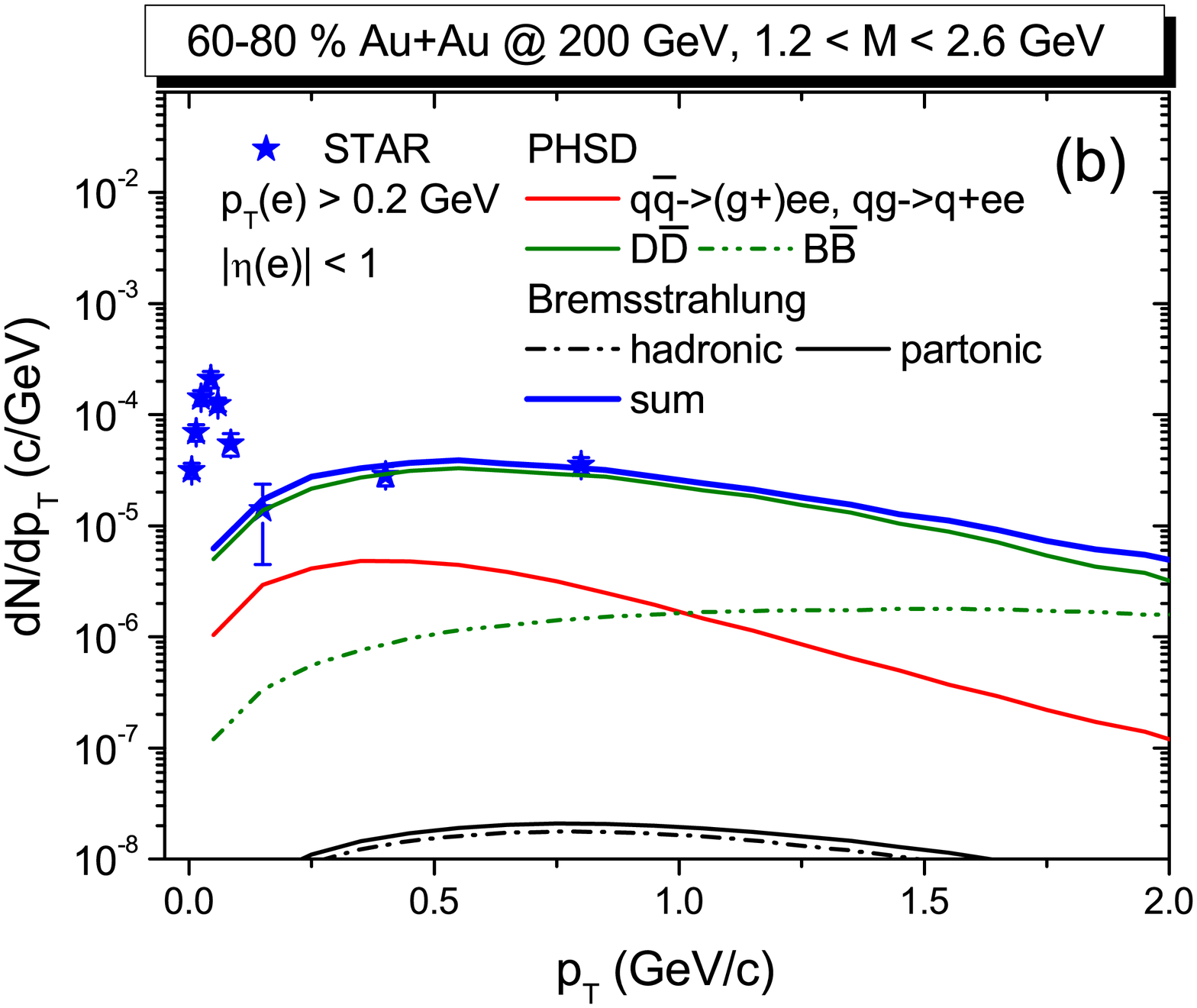}}
\caption{Transverse momentum spectra of (a) low-mass and (b)
intermediate-mass dileptons with the individual contributions shown
additionally in 60-80 \% central Au+Au collisions at $\sqrt{s_{\rm
NN}}=$ 200 GeV. The experimental data are taken from
Ref.~\cite{yang2018}.} \label{peripheral}
\end{figure}

\begin{figure} [h!]
\centerline{
\includegraphics[width=8.2 cm]{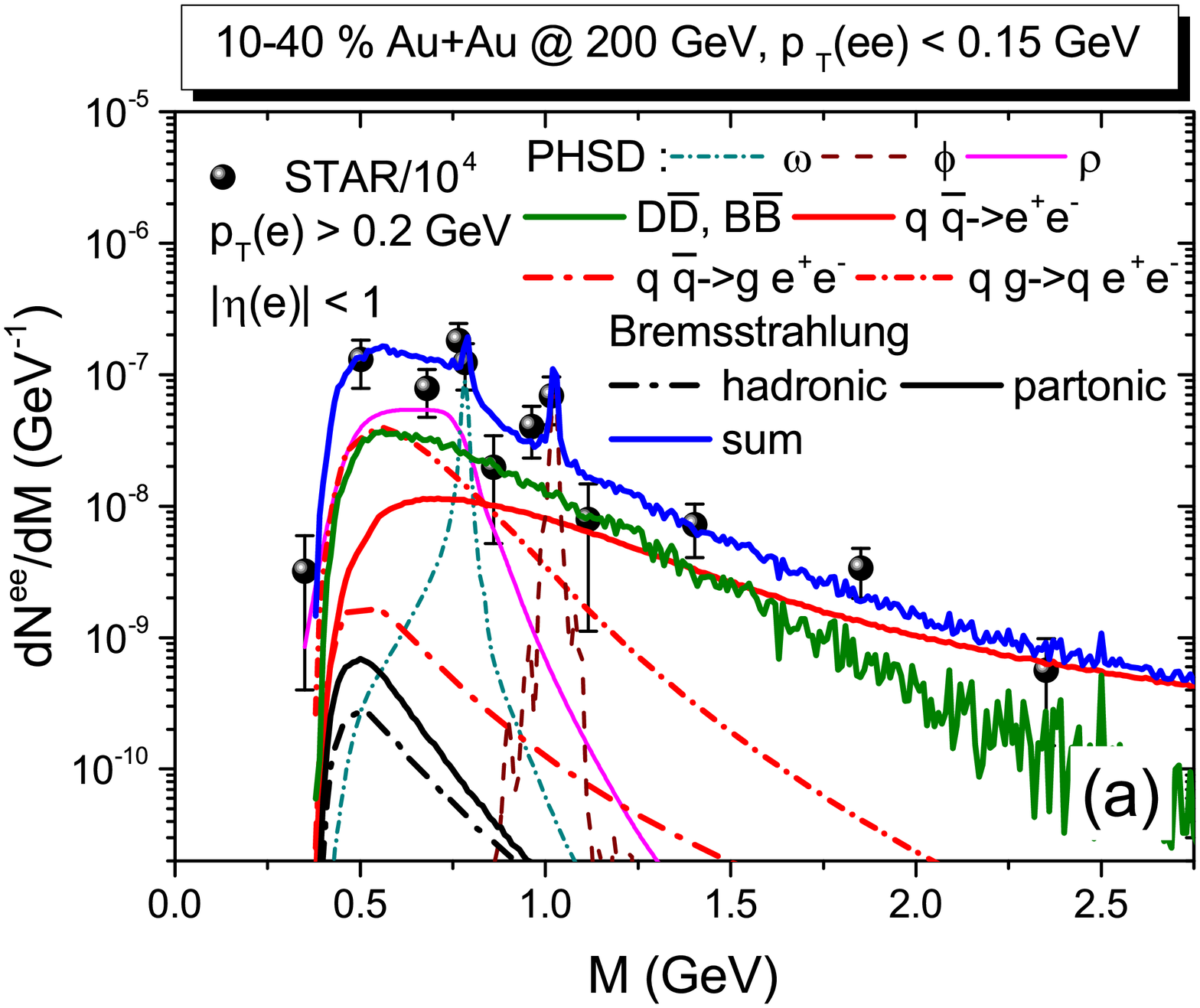}}
\centerline{
\includegraphics[width=8.2 cm]{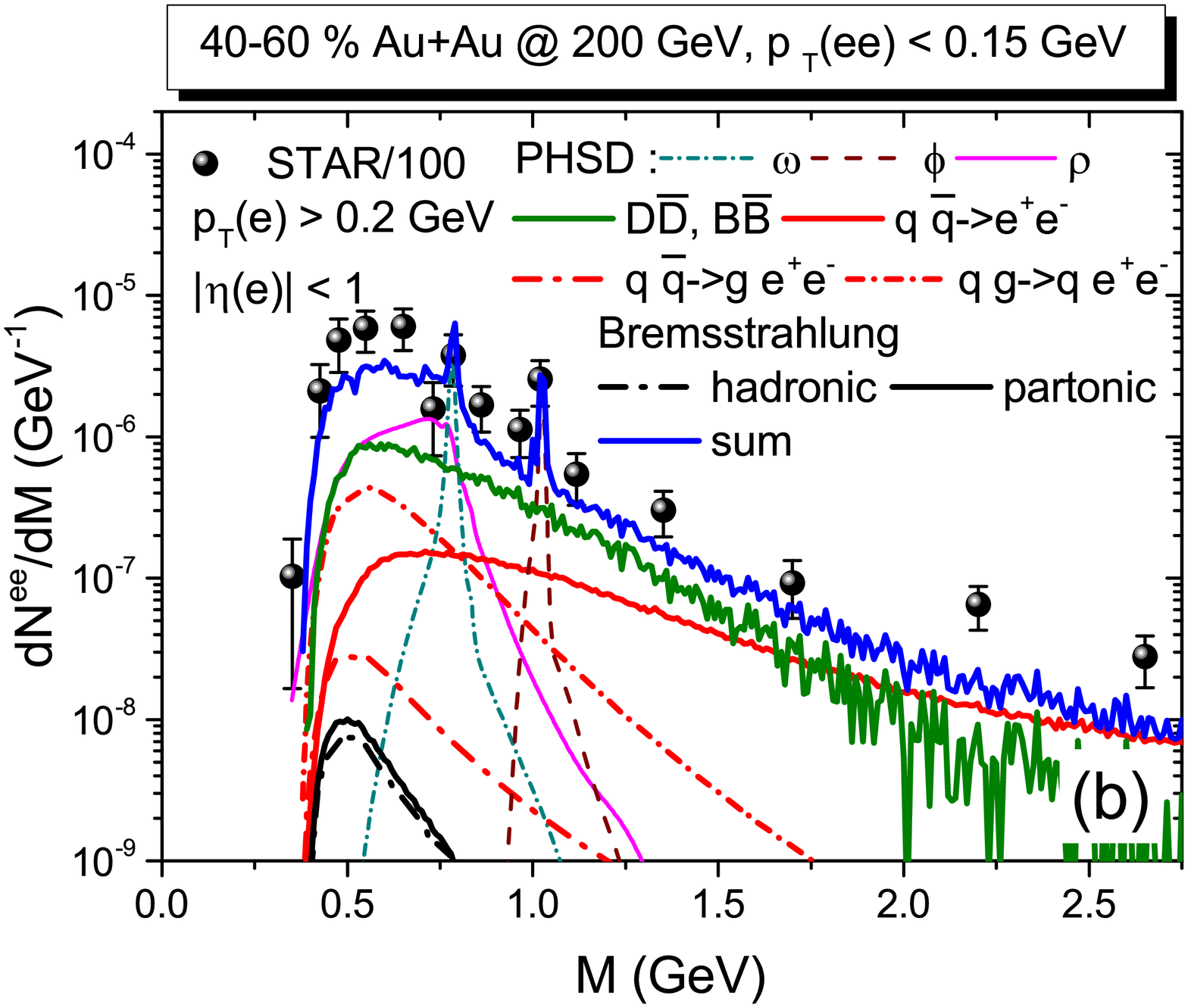}}
\centerline{
\includegraphics[width=8.2 cm]{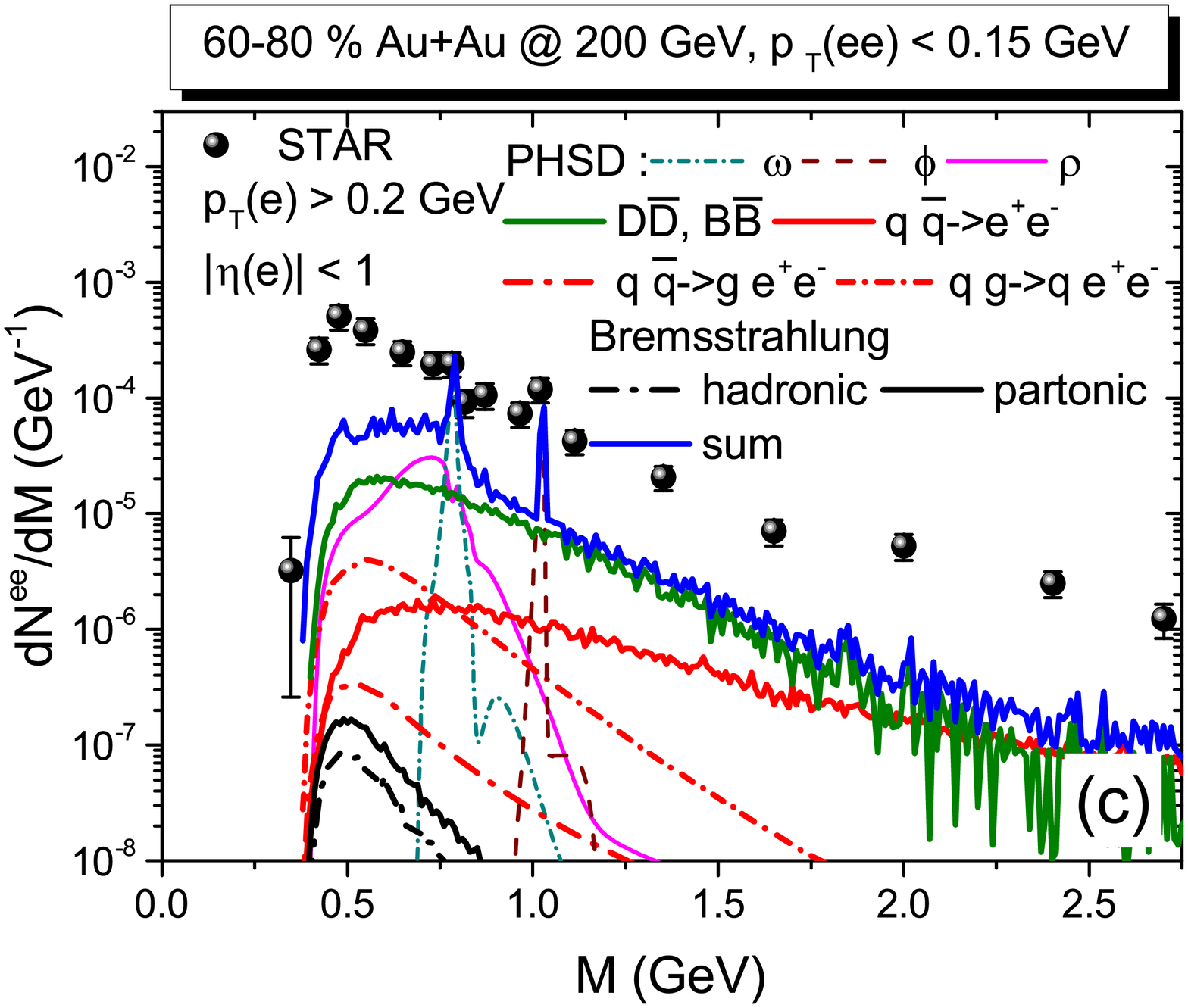}}
\caption{Invariant mass spectra of dileptons with small transverse
momentum ($p_T < 0.15$ GeV/c) in (a) 10-40, (b) 40-60, and (c) 60-80
\% central Au+Au collisions at $\sqrt{s_{\rm NN}}=$ 200 GeV. The
experimental data are taken from Ref.~\cite{yang2018}.}
\label{low-pt}
\end{figure}

In order to provide further information, we show in Fig.
\ref{peripheral}  all contributions to the transverse momentum
spectra of low-mass and intermediate-mass dileptons in 60-80 \%
central collisions. As in Fig. \ref{rhic}, the low-mass dilepton
sector has contributions from various hadronic and partonic channels
and the most dominant contributions are from $D\bar{D}$ pairs and
$\rho$-meson decays. On the other hand, in the intermediate-mass
dilepton sector the contribution from  $D\bar{D}$ pairs and partonic
interactions are dominant with some background from $B\bar{B}$
pairs. As mentioned in the previous section, there are three kinds
of nuclear modifications on dileptons in heavy-ion collisions, but
none of them can explain the enhancement of dileptons at low
transverse momentum. We recall that the dileptons from heavy flavor
pairs are not visibly modified in very peripheral collisions due to
the low amount of rescattering as shown in Fig. \ref{ddbar}. Also
the contributions from $\rho$-meson decays  or partonic
interactions are subdominant.
Furthermore, the dilepton Bremsstrahlung is peaked at low transverse momentum only for very small invariant mass, $M\rightarrow 0$, while in the two mass regions of interest the $p_T$ spectra are broad and show no indication
for a low $p_T$ peak.

Fig. \ref{low-pt}, furthermore, shows the invariant mass spectra of
dileptons with small transverse momentum ($p_T < 0.15$ GeV/c) in
10-40, 40-60, and 60-80 \% central Au+Au collisions at $\sqrt{s_{\rm
NN}}=$ 200 GeV in comparison to the experimental data from
Ref.~\cite{yang2018}. The PHSD can reproduce the experimental data
in 10-40 \% central collisions very well, but begins to deviate
slightly from the data in 40-60 \% central collisions; the deviation
becomes pronounced for 60-80 \% central collisions, which is consistent
with Fig. \ref{peripheral}, and  implies that the anomalous
enhancement of dileptons at low transverse momentum is only small or moderate
in central collisions. If we assume that the
dilepton spectrum is the same at low transverse momentum (in Fig.
\ref{spec1}) regardless of centrality, then the anomalous source is
quite strong in 60-80 \% central collisions, less strong in 40-60 \%
central collisions, and hardly seen in 10-40 \% central collisions.
Furthermore, since the $p_T$ range is very small we conclude that
the transverse mass distribution from the anomalous source is almost
the same as from hadronic and partonic contributions in central
collisions.

The other point is that differences between the experimental data
and the PHSD results in 40-60 and 60-80 \% central collisions do
practically not depend on the invariant mass of the dileptons but
are rather constant in magnitude. For example, the yield of low-mass
dielectrons ($0.4 < M < 0.79$ GeV) and that of intermediate-mass
dielectrons ($1.2 < M < 2.6$ GeV) from the experimental data in
60-80 \% central collisions are alike but about ten times larger
than those from the PHSD. These findings are hard to reconcile.

In summarizing we have addressed the low $p_T$ enhancement of
dileptons from peripheral heavy-ion collisions where the
experimental data show a large anomalous source regardless of the
dilepton invariant mass. We have employed  the PHSD transport
approach to describe the transverse momentum spectra of dileptons in
relativistic heavy-ion collisions which incorporates  three
in-medium effects in heavy-ion collisions: i) The spectral functions
of vector meson broaden in nuclear matter, ii) the correlation of
heavy-flavor pairs is modified by partonic and hadronic
interactions, and iii) there are sizeable contributions from
partonic interactions which do not exist in hadronic cocktails.
Taking all matter effects into account, the PHSD reproduces the
experimental data for dileptons down to $p_T \approx$ 0.15 GeV/c at
all centralities, however,  underestimates the data below $p_T
\approx$ 0.15 GeV/c in very peripheral collisions.
In extension of previous studies we have incorporated the production
of dilepton pairs by hadronic and partonic bremsstrahlung processes - as suggested early in Refs.
\cite{add1,add2,add3,add4,add5,add6} - employing the soft-photon approximation for an estimate.
We find that these radiative corrections are by far subleading and - in the invariant mass regions of interest -
do not peak at low $p_T$. Accordingly, the
large enhancement of dileptons at low transverse momentum in
peripheral heavy-ion collisions is still an open question and the
solution of the puzzle is beyond standard microscopic models that
have shown to be compatible with dilepton data from heavy-ion
collisions in the range from SIS to LHC energies \cite{PHSDreview}.
\\
\\
The authors acknowledge inspiring discussions with
J. Butterworth, F. Geurts and C. Yang.
This work was supported by the LOEWE center "HIC for FAIR",  the
HGS-HIRe for FAIR and the COST Action THOR, CA15213. Furthermore, PM
and EB acknowledge support by DFG through the grant CRC-TR 211
'Strong-interaction matter under extreme conditions'. The
computational resources have been provided by the LOEWE-CSC.

\end{document}